\documentstyle[aps]{revtex}

\newcommand{\bc}{\begin{center}}
\newcommand{\ec}{\end{center}}
\newcommand{\be}{\begin{equation}}
\newcommand{\ee}{\end{equation}}
\newcommand{\bq}{\begin{eqnarray}}
\newcommand{\eq}{\end{eqnarray}}

\begin{document}
\noindent

\vspace*{.1in}

\begin{center}
{\Large \bf Interface dynamics for layered structures}

\vspace*{.1in}
Takao Ohta$^{* \dagger}$ and David Jasnow$^{\dagger}$ \\
{\sl${}^{\dagger}$ Department of Physics and Astronomy, 
University of Pittsburgh }\\
{\sl Pittsburgh, PA 15260}\\
{\sl${}^{*}$ Department of Physics, Ochanomizu University, Tokyo 112 JAPAN}

{\sl  \today}

\end{center}


\begin{abstract}

\noindent
We investigate dynamics of large scale 
and slow deformations of layered structures.
Starting from the respective model equations for a non-conserved system,
a conserved system and a binary fluid, we derive the
interface equations which are a coupled 
set of equations for deformations
of the boundaries of each domain.  A further reduction of the
degrees of freedom is possible for a non--conserved system 
such that internal motion of 
each domain is adiabatically eliminated. 
The resulting equation of motion contains only the
displacement of the center of gravity of domains, 
which is equivalent to the phase
variable of a periodic structure. 
Thus our formulation automatically includes the phase
dynamics of layered structures.
In a conserved system and a binary fluid, however, the internal
motion of domains turns out to be a slow variable in the long
wavelength limit because of concentration conservation. Therefore
a reduced description only involving the phase variable is not 
generally justified.
\end{abstract}
\vspace*{.1in}

PACS number(s): 68.10.-m, 68.60.Dv, 68.15.+e

\pagebreak

\section{Introduction}
Layered structures appear in various systems in nature.
Smectic liquid crystals \cite{dG}, mesophases in a microphase separation 
of block copolymers \cite{BF}, magnetic bubbles 
in magnetic thin films \cite{SA} and condensed phases of Langmuir
monolayers \cite{KNO}, \cite{MOH}
are typical examples in thermal equilibrium.

A layered structure can also be observed in pattern formation
far from equilibrium \cite{NATO}. Turing was the first to predict a stable
periodic structure in a reaction-diffusion system. Recently, 
a Turing instability has been observed experimentally in a chemical
reaction in a gel \cite{SW}. Other well known examples are 
roll patterns in Rayleigh-Benard convection and Williams domains
in the electrohydrodynamic convection of liquid crystals.

In order to study dynamics of these structures, a coarse-grained
description is often employed.  Suppose that there is a periodic 
layer with period $\ell$ in x-direction. 
This is expressed by a periodic 
function $X(x+\phi)=X(x+\ell+\phi)$ where 
$X$ is a physical property and $\phi$ is a phase and is arbitrary
in a uniform extended system. 
When the periodic structure is slightly deformed,
its large scale slow motion can be described by allowing $\phi$ to
depend on $x$, $y$, and $t$. 
Since $\phi$ is a neutral mode, it is one of the slowest
variables in the long wavelength limit. The theory based on this idea,
which is called phase dynamics, was developed in 
pattern dynamics far from equilibrium \cite{PM}, \cite{KURA}. 

An essentially similar method can be applied to layered structures
in equilibrium \cite{CL}. One can express the free energy increase due to
deformation in terms of the local displacement field as in continuum
elastic theory. The local displacement field corresponds to the phase
variable and the dynamic equation for it can be written down by using the
free energy functional.

Many of the theories assume a sinusoidal function 
(and only a few higher harmonics) of $X(x)$ in the derivation of
the phase equation. This is valid when the system is in the vicinity
of a bifurcation (critical) point where a uniform state loses stability and
a periodic structure appears supercritically. However, when the function $X$
changes abruptly at the interface separating two adjacent layers, we need to
develop an alternative method for dynamics. Phase dynamics is 
still meaningful in this case for long wavelength deformations, 
but the derivation of the phase equation is not possible 
using available theories.

In this paper, we address the above problem. Some years ago, Kawasaki and
one of the authors \cite{KO1} derived a 
free energy functional of general periodic 
structures in terms 
of the displacement fields without assuming the slow variation of $X$.
Recently a similar analysis has been
applied to a model specific to Langmuir monolayers \cite{DS}.
Here, on the other hand, we are concerned with dynamics. 
The static aspects can be obtained as a byproduct 
by taking the equilibrium limit. It is emphasized that 
our theory does not necessarily rely on the existence 
of a free energy functional or a Lyapunov functional.
As a consequence, we can deal with the systems
both near equilibrium and far from equilibrium in a unified way. 

Our strategy is to derive the interface equation of motion for each
domain boundary. Suppose that $X$ varies within one period as 
$X=1$ for $0<x<2w$ and $X=-1$ for $2w<x<\ell$ where $w(<\ell/2)$ is a constant.
Thus there are interfaces at $x=0$ and $x=2w$ 
bounding a domain where $X=1$.
We assume the interface width is infinitesimal compared to $w$ and $\ell$.
It is generally expected that 
these two interfaces are coupled to each other
as well as to those in other domains. 
We can derive the phase equation of motion from the interface equation
by retaining only the average displacement of two interfaces at each domain.
The other degree of freedom which makes the center gravity of a domain 
time--independent represents the internal motion of domains. In some
situations, 
internal deformations of each domain are relevant to the stability. 
The present
theory covers such a domain instability. 

The phase variable describes the Goldstone mode associated with the 
translational symmetry breaking due to a periodic structure in an
extended uniform system, and hence it is a slow variable in the long wavelength
limit. In the terminology of lattice vibrations this corresponds to
an accoustic mode. On the other hand, the internal motion of domains is
similar to an optical mode and has a 
finite energy in the limit of long wavelength
deformation. As will be shown below, 
this gives rise to a finite relaxation rate of the internal
mode for small wavenumbers in a non--conserved case. 
Therefore one may eliminate adiabatically the internal mode. 
However, as we will show, the situation is entirely different 
in a conserved case and a binary fluid.
When the interface width is infinitesimal, and there is no
mass flux through the system boundary, concentration conservation
is equivalent to a conservation of total area 
(volume in three dimensions) of domains.
Therefore, the internal mode should also be a slow variable in the
long wavelength limit. Our main concern in this paper 
is to clarify how the phase dynamics
are affected by the extra slow mode originating from conservation.

We start with a time--dependent Ginzburg--Landau (TDGL) type equation
describing the system at the coarse--grained level. 
A general method is available to
derive the interface equation of motion \cite{KO2} from the TDGL equation. 
In binary alloys or in crystal growth, on the other hand,
it is also well--known that the same equation
can be obtained by starting at the macroscopic level 
with the simple diffusion equation for
the minority atoms and employing the Gibbs--Thomson 
relation as the boundary condition at an infinitely sharp interface
\cite{INT}. Thus one may wonder
why the TDGL approach is necessary. There are, of course, several reasons.
First of all, most simulations of spinodal decomposition 
and other phase separation phenomena in binary systems
use TDGL equations since for many purposes 
it is more efficient than the macroscopic approach.
Hence the interfacial analysis based on the TDGL equation 
is necessary for a complete understanding of the simulation results. Secondly, 
the interfacial method is valid beyond the simple diffusion equation
which has been used only for macrophase separation and for crystal growth. 
As mentioned above, we will show that 
our method can be useful for both thermal equilibrium structures and 
dissipative structures out of equilibrium.

The organization of this paper is as follows.
In section 2, we make a general argument on layered structures 
in equilibrium. In section 3, we derive the interface equation of motion
for a periodic structure in a non-conserved case. 
In section 4, we verify that the vanishing of one of 
the diffusion constants in the phase equation is equivalent to
free energy minimization with respect to the spatial period.
The method is
extended in section 5 to a conserved system. In section 6, we treat 
a layered structure in binary fluids. 
The results obtained are discussed and summarized in section 7.
Some of the details of the method used in section 3 are
described in the Appendix.

\section{Periodic structures}
\label{sec:peri}
A typical example of a layered structure in a system
with a non-conserved order parameter is a magnetic thin film \cite{GD}.
Experiments \cite{SEUL} show stripe and bubble structures of magnetic domains
depending on the magnitude of the external magnetic field normal to
the film. The free energy of this system can be written in terms of
the local order parameter $u(\vec{r})$ as
\be
F=F_{GL} +F_{LR}
\label{eq:free}
\ee
where
\be
F_{GL}=\int d\vec{r}[\frac{\epsilon^2}{2}(\vec {\nabla} u)^2 + W(u)-hu]
\label{eq:fgl}
\ee
and
\be
F_{LR}=\frac{\alpha}{2} \int d\vec{r}\int d\vec{r}'G(\vec{r},\vec{r}')
[u(\vec{r})-c] [u(\vec{r}')-c]
\label{eq:flr}
\ee
Note that the free energy is generally divided into two parts. 
$F_{GL}$ is the short ranged part and is the usual 
Ginzburg-Landau form with $W(u)$ a double well potential
in the ordered state and $h$ the external field. 
The parameter $\epsilon$, which
is a measure of an interfacial width,
is assumed to be sufficiently small. The other part $F_{LR}$ contains the 
long range interaction, where $\alpha$ is the interaction strength and 
$c$ is a constant. The form of the kernel 
$G(\vec{r},\vec{r}')$ depends on the system considered. 
For instance, it is given in a thin magnetic film of a strongly uniaxial
material by \cite{GD}
\be
G(\vec{r},\vec{r}')=\int d\vec{q} \frac{4\pi}{q}[1- \exp(-qD)]
\exp(i\vec{q}\cdot(\vec{r}-\vec{r}'))
\label{eq:mag}
\ee
where the integral over $\vec{q}$ is taken in a two-dimensional
space parallel to the film surface.
Note that the order parameter $u$ in this case is the magnetization
normal to the film which has thickness $D$.
The spatial period of a modulated structure is larger than $D$,
and the interface width $\epsilon$ is much smaller than $D$, i.e.,
$\epsilon<<D<<\ell$.
Probably the simplest choice of $G(\vec{r},\vec{r}')$ which provides us with
a periodic structure in equilibrium is the Coulomb Green function
\be 
-\nabla^2G_c(\vec{r},\vec{r}')=\delta (\vec{r}-\vec{r}')
\label{eq:green}
\ee
This type of long range interaction appears 
in the theory of phase separation of 
block copolymers \cite{OK1}. The constant $c$ in (\ref{eq:flr}) is equated 
to the spatial average of $u$ to avoid a divergence
in the long range interaction.

In terms of Fourier components, the sum of the short range 
and the long range
interactions contains the quadratic part
\be 
\frac{1}{2}\Gamma(q)u_{\vec{q}}u_{-\vec{q}}
\label{eq:min1}
\ee
where
\be 
\Gamma(q)=\epsilon^2 q^2+\alpha G_q -\tau
\label{eq:min2}
\ee
with $G_q$ the Fourier transform of $G(\vec{r}-\vec{r}')$.
The constant $\tau$ comes from the bilinear term in the local part $W(u)$.

In order for a stable periodic structure to exist, we assume that
$\Gamma$ has a minimum for a finite value of $q$ and that the free energy
(\ref{eq:free}) takes the lowest value for such a modulation. Hereafter,
to be specific, we 
consider the case where $G(\vec{r},\vec{r}')$ is given by the
screened Coulomb interaction
\be 
-(\nabla^2-\kappa^2)G(\vec{r},\vec{r}')=\delta (\vec{r}-\vec{r}')
\label{eq:greenkappa}
\ee
where $1/\kappa$ is the screening length which is assumed to be 
much larger than $\epsilon$.
In this sense, the interaction $G(\vec{r}, \vec{r}')$
is still long ranged. 
The interfacial theory for other forms of 
$G(\vec{r},\vec{r}')$ can be constructed similarly; in particular, the
pure Coulomb case follows by taking the $\kappa \rightarrow 0$ limit
of the interface equations.

\section{Non-conserved system}
\label{sec:non}
The time-evolution equation of 
a non-conserved order parameter $u$ is determined by the free energy 
functional (\ref{eq:free}) according to
\be
\frac{\partial u}{\partial t} 
=-\frac{\delta F}{\delta u} 
\label{eq:tdgl}
\ee
It is noted that the right hand side contains the nonlocal long range
interaction. When we derive the interface equation of motion
for layers, it is more convenient to express it in a local form.
To this end, we write (\ref{eq:tdgl}) 
in the form of a coupled set of equations as
\be
\tau_1\frac{\partial u(\vec{r})}{\partial t} 
=\epsilon^2\nabla^2 u + f(u)-v+h 
\label{eq:bvp1}
\ee
\be
\tau_2\frac{\partial v(\vec{r})}{\partial t} 
=\nabla^2 v -\kappa^2 v+ \alpha u
\label{eq:bvp2}
\ee
where we have introduced two parameters $\tau_1$ and $\tau_2$ and $f(u)$, 
defined by
\be
f(u)=-\frac{dW}{du}
\label{eq:fu}
\ee
The variable $v$ in (\ref{eq:bvp2}) is an auxilliary field and
has no physical meaning for the model (\ref{eq:tdgl}).
However the set of equations (\ref{eq:bvp1}) and (\ref{eq:bvp2})
appears in pattern formation far from equilibrium as will be mentioned below.
Without loss of generality, we may put the diffusion constant for $v$ 
equal to unity.
In the limit $\tau_2 \rightarrow 0$, eq.(\ref{eq:bvp2}) can be solved for $v$ by
using $G(\vec{r},\vec{r}')$ defined by (\ref{eq:greenkappa}). Substituting
it into
eq. (\ref{eq:bvp1}) we recover eq. (\ref{eq:tdgl}).

For a general form of the kernel $G(\vec{r},\vec{r}')$ in the long
range interaction, we may replace $\nabla^2 v-\kappa^2 v$ by 
$\int d\vec{r}'\Gamma(\vec{r},\vec{r}')v(\vec{r}')$ 
where $\Gamma(\vec{r},\vec{r}')$ is defined through the relation
\be
\int d\vec{r}"G(\vec{r},\vec{r}")\Gamma(\vec{r"},\vec{r}')
=\delta(\vec{r}-\vec{r}")
\label{eq:kernel}
\ee

It is remarked that eqs. (\ref{eq:bvp1}) and (\ref{eq:bvp2}) are model equations
for pattern dynamics far from equilibrium. For instance, the
Belousov-Zhabotinski
reaction in the excitable regime can be described, in its simplest
approximation, 
by eqs. (\ref{eq:bvp1}) and (\ref{eq:bvp2}). In fact this set of equations 
has been extensively studied during the past two decades 
by applied mathematicians \cite{TK}, \cite{KO}. 
It is well known that eqs. (\ref{eq:bvp1}) and (\ref{eq:bvp2}) undergo a
Turing instability when $\epsilon$ is small so that a periodic structure
can be formed. To our knowledge, however, an interfacial approach has not been
applied to a periodic solution except for ref. \cite{OIT}.

We solve eqs. (\ref{eq:bvp1}) and (\ref{eq:bvp2}) in the limit $\epsilon
\rightarrow 0$ to derive 
the interface equation of motion. We assume that 
when the diffusion terms are absent, these equations have two stable uniform
solutions $(u_1, v_1)$ and $(u_2, v_2)$. As shown in Fig. 1, we consider 
a layer of domains in two dimensions where the regions $u=u_1$ and $u=u_2$ 
are arrayed alternatively. We assume the width of a domain in which $u=u_1$ 
is equal
to $2w$ and the period of the layer is $\ell$. 

Now suppose these striped domains are gently curved as in Fig. 1. We introduce
the deformations $\zeta_n^+(y, t)$ ($\zeta_n^-(y, t)$) around the equilibrium 
position of the right (left) interface in the n-th
domain in which $u=u_1$. Our aim is to derive the linearized equation of 
motion for $\zeta_n^{\pm}$.
For concreteness, we hereafter 
assume the following 
form of $f(u)$
\be
f(u)=\frac{1}{2}u(1-u^2)
\label{eq:fuu}
\ee

As was mentioned above, the parameter $\epsilon$ is a measure of an interface
width. 
Suppose that we are concerned with the motion of a specific interface.
In the limit $\epsilon \rightarrow 0$, the density $u$ varies rapidly 
through the interface, but the spatial variation of $v$ is much slower.
Therefore we may replace $v$ in eq. (\ref{eq:bvp1})
by its value at the interface, which is denoted by $v_I$ 
and which may depend on time and position on the interface. 
Thus eq. (\ref{eq:bvp1}) can be written as
\be
\tau_1\frac{\partial u(\vec{r})}{\partial t} 
=\epsilon^2\nabla^2 u + f(u)-v_I+h 
\label{eq:bvp1I}
\ee
This is just the TDGL equation with the "external
field", $v_I-h$, and hence the interface equation of motion 
for this system is readily obtained: 
\be
\tau_1 V=\epsilon^2H - 3\epsilon(v_I-h)
\label{eq:kaimen}
\ee
where, for simplicity, we have assumed that the absolute value of 
$v_I-h$ is sufficiently small so that the gap of the density $u$ at the
interface is  approximately equal to 2. 
$V$ is the normal component of the interface velocity and $H$ the mean 
curvature. 
These are defined as follows. First let us introduce a field variable 
$\psi(\vec{r}, t)$ which is positive (negative) in the region $u>0$ ($u<0$)
and $\psi(\vec{r}, t)=0$ specifies the interface configuration. The velocity $V$
and the mean curvature $H$ are expressed respectively as
\be
V=\frac{1}{|\vec{\nabla}\psi|}\frac{\partial \psi}{\partial t}
\label{eq:vel}
\ee
\be
H=\vec{\nabla}\cdot \vec{n}
\label{eq:mean}
\ee
with
\be
\vec{n}=\frac{\vec{\nabla}\psi}{|\vec{\nabla}\psi|}
\label{eq:unit}
\ee
Note that $\vec{n}$ is the {\em inward} normal of a $u>0$ domain.

From the above definition of $\psi$, deformations $\zeta^{\pm}$ 
may be written as
$\psi=-x+\zeta^+(y, t)$ and $\psi=x-\zeta^-(y, t)$. Substituting these into 
(\ref{eq:vel}), (\ref{eq:mean}) and (\ref{eq:kaimen}),
we obtain up to linear order in
$\zeta^{\pm}$ 
\be
\tau_1\partial \zeta_n^{\pm}/\partial t=
\epsilon^2\partial^2 \zeta_n^{\pm}/\partial y^2\mp 3\epsilon (v_I-h)
\label{eq:zeta}
\ee
The last term indicates that a domain where $u>0$ shrinks as $v_I-h$
is increased.

The remaining problem is to determine $v_I$ as a functional
of $\zeta_n^{\pm}$ so that eq. (\ref{eq:zeta}) becomes a closed set of 
equations. In what follows, we need to solve eq. (\ref{eq:bvp2}) for a given 
set of $\zeta_n^{\pm}$. This was carried out in a previous paper \cite{OIT} for 
general values of $\tau_2$ to study a dynamical instability of layered domains.
Here we consider only the limit $\tau_2 \rightarrow 0$. 
In this limit, the solution
is given by
\be
v(\vec{r},t)=\alpha\int d\vec{r}'G(\vec{r},\vec{r}')u(\vec{r}',t)
\label{eq:vrt}
\ee
When the absolute value of 
$v_I-h$ is sufficiently small as we have assumed,
the uniform solutions $u_1$ and $u_2$ may be approximated as
$u_1=1$ and $u_2=-1$. 
Actually only the difference $u_1-u_2$ enters in the theory given below.
The correction is however, of order $(v_I-h)^2$, i.e., 
$u_1-u_2 = 2+O((v_I-h)^2)$ and hence does not affect
the first order correction in (\ref{eq:kaimen}).
Therefore eq. (\ref{eq:vrt}) can be written as
\begin{eqnarray}
v(\vec{r},t)&=&\alpha\int \frac{d\vec{q}}{(2\pi)^2}\int dy'
\sum_{m}[\int_{b_m^-}^{b_m^+}dx'\frac{1}{q^2+\kappa^2}
\exp(i\vec{q}\cdot(\vec{r}-\vec{r}'))\nonumber\\
{}&-&\int_{b_m^+}^{b_{m+1}^-}dx'\frac{1}{q^2+\kappa^2}
\exp(i\vec{q}\cdot(\vec{r}-\vec{r}'))]
\label{eq:vrtm}
\end{eqnarray}
where we have defined
\be
b_m^{\pm}=m\ell \pm w+\zeta_m^{\pm}
\label{eq:vrtpm}
\ee
The value $v_I$ in the n-th domain, which generally takes a different value
at the right and the left interfaces, is given by
\be
v_I^{\pm}=v(b_n^{\pm}, y, t)
\label{eq:vIn}
\ee

Now we calculate $v_I$ to first order in $\zeta^{\pm}$. 
Some further details are provided in the Appendix.
The zeroth order solution is given from eq. (\ref{eq:a2}) in the Appendix by
\be
v_I^{(0)\pm}=\frac{\alpha}{\ell}\sum_{Q}\frac{2}{Q(Q^2+\kappa^2)}\sin 2Qw
\label{eq:vI0}
\ee
where reciprocal lattice vectors are given by
$Q=2\pi n/\ell$ with $n$ integers ($n=0, \pm1, \pm2, ...$).
The first order solution is given from eq. (\ref{eq:a3}) in the Appendix by
\begin{eqnarray}
v_I^{(1)+}&=&-2\alpha \zeta_n^+ (y)\frac{1}{\ell}
\sum_{Q}\frac{1}{Q^2+\kappa^2}(1-\cos 2Qw)
\nonumber\\
{}&+&2\alpha \int \frac{dk}{2\pi}e^{in\ell k}
\int \frac{dp}{2\pi} \frac{1}{\ell}
\sum_{Q}\frac{e^{ipy}}{(k-Q)^2 +p^2+\kappa^2}\nonumber\\
{}&\times &[\zeta_{k,p}^+-e^{-2iQw}\zeta_{k,p}^-]
\label{eq:vI1p}
\end{eqnarray}

\begin{eqnarray}
v_I^{(1)-}&=&2\alpha \zeta_n^- (y)\frac{1}{\ell}
\sum_{Q}\frac{1}{Q^2+\kappa^2}(1-\cos 2Qw)
\nonumber\\
{}&+&2\alpha \int \frac{dk}{2\pi}e^{in\ell k}
\int \frac{dp}{2\pi} \frac{1}{\ell}
\sum_{Q}\frac{e^{ipy}}{(k-Q)^2 +p^2+\kappa^2}\nonumber\\
{}&\times & [e^{2iQw}\zeta_{k,p}^+-\zeta_{k,p}^-]
\label{eq:vI1m}
\end{eqnarray}
In the derivation of eqs. (\ref{eq:vI1p})
and (\ref{eq:vI1m})
we have introduced the Fourier transform
\be
\zeta _n^\pm (y)=\int \frac{dk}{2\pi}\int \frac{dp}{2\pi}
\zeta _{k,p}^\pm \exp [ik(n\ell \pm w) +iyp]
\label{eq:zetaK}
\ee
and used the Poisson formula
\be
\sum_{n=-\infty}^{\infty}\exp (ikn\ell )=\frac{1}{\ell}\sum_{Q}\delta (k-Q)
\label{eq:poissonf}
\ee
Substituting these into eq. (\ref{eq:zeta}) 
we obtain the interface equation of motion.
The zero-th order solution is particularly important, which reads
\be
h=\frac{2\alpha}{\ell}\sum_{Q}\frac{1}{Q(Q^2+\kappa^2)}\sin 2Qw
\label{eq:kvI0}
\ee
This specifies the ratio $w/\ell$ of these unknown parameters. 
It is readily verified that when $h=0$, $w/\ell = 1/4$, as it should be
for the symmetric case. Equation (\ref{eq:kvI0}) is equivalent to
$\partial \zeta_n^{\pm}/\partial t =0$ without deformations. Hence
this is an equilibrium condition for a nonconserved system.

From the first order equation, we obtain the coupled set of linear
equations for $\zeta^{\pm}$:
\begin{eqnarray}
\tau_1\frac{d}{dt}\zeta_{k,p}^{+}&=&-[\epsilon^2p^2-\frac{6\epsilon\alpha}{\ell}
\sum_{Q}\frac{1}{Q^2+\kappa^2}(1-\cos 2Qw)]\zeta_{k,p}^{+}\nonumber\\
{}&+&\frac{6\epsilon\alpha}{\ell}\sum_{Q}\frac{1}{(k-Q)^2+p^2+\kappa^2}
[-\zeta_{k,p}^+ +e^{-2iQw}\zeta_{k,p}^-]
\label{eq:eqzetap}
\end{eqnarray}

\begin{eqnarray}
\tau_1\frac{d}{dt}\zeta_{k,p}^{-}&=&-[\epsilon^2p^2-\frac{6\epsilon\alpha}{\ell}
\sum_{Q}\frac{1}{Q^2+\kappa^2}(1-\cos 2Qw)]\zeta_{k,p}^{-}\nonumber\\
{}&+&\frac{6\epsilon\alpha}{\ell}\sum_{Q}\frac{1}{(k-Q)^2+p^2+\kappa^2}
[e^{2iQw}\zeta_{k,p}^+ -\zeta_{k,p}^-]
\label{eq:eqzetam}
\end{eqnarray}
In order to analyze the motion of long wavelength deformations, we make the
following transformation
\be
\zeta_{n}^{(1)}(y)=\frac{1}{2}(\zeta_{n}^+(y) +\zeta_{n}^-(y))
\label{eq:zeta1}
\ee

\be
\zeta_{n}^{(2)}(y)=\frac{1}{2}(\zeta_{n}^+(y) -\zeta_{n}^-(y))
\label{eq:zeta2}
\ee
In real space, $\zeta_{n}^{(1)}$ describes the deviation of
the center of gravity of each domain while $\zeta_{n}^{(2)}$
is the internal deformation of a domain. 
Note that $\zeta_{n}^{(1)}\ne 0$ but $\zeta_{n}^{(2)}= 0$
for a uniform
translation of a domain. Thus $\zeta_{n}^{(1)}$ is a phase variable.

Since we are concerned with large scale deformations of a modulated
structure, we may apply a gradient expansion to 
eqs. (\ref{eq:eqzetap}) and (\ref{eq:eqzetam}).
That is, we expand the coefficients of these equations in terms of $k$ and $p$.
Retaining the lowest nontrivial order, we obtain 
from eqs. (\ref{eq:eqzetap}) and (\ref{eq:eqzetam})
\begin{eqnarray}
\tau_1\frac{d}{dt}\zeta_{k,p}^{(1)}&=&-[\epsilon^2 p^2+
\frac{6\epsilon\alpha}{\ell}(3k^2-p^2)
\sum_{Q}\frac{1}{(Q^2+\kappa^2)^2}(1-\cos 2Qw)]\zeta_{k,p}^{(1)}\nonumber\\
{}&+&\frac{12i\epsilon\alpha}{\ell}k\zeta_{k,p}^{(2)}
\sum_{Q}\frac{Q}{(Q^2+\kappa^2)^2}\sin 2Qw\nonumber\\
{}&-&\frac{6\epsilon\alpha}{\ell}\sum_{Q}\frac{1}{(Q^2+\kappa^2)^3}(1-\cos
2Qw)p^4\zeta_{k,p}^{(1)}
\label{eq:eqzeta1}
\end{eqnarray}

\begin{eqnarray}
\tau_1\frac{d}{dt}\zeta_{k,p}^{(2)}&=&-[\epsilon^2p^2+
\frac{12\epsilon\alpha}{\ell}\sum_{Q}\frac{1}{Q^2+\kappa^2}\cos
2Qw]\zeta_{k,p}^{(2)}\nonumber\\
{}&-&\frac{6\epsilon\alpha}{\ell}(3k^2-p^2)
\sum_{Q}\frac{1}{(Q^2+\kappa^2)^2}(1+\cos 2Qw)\zeta_{k,p}^{(2)}\nonumber\\
{}&-&\frac{12i\epsilon\alpha}{\ell}k\zeta_{k,p}^{(1)}\sum_{Q}\frac{Q}{(Q^2+\
\kappa^2)^2}\sin 2Qw
\label{eq:eqzeta2}
\end{eqnarray}
Note that the right hand side of eq. (\ref{eq:eqzeta1}) vanishes in the limit
$k, p \rightarrow 0$ whereas that of (\ref{eq:eqzeta2}) 
remains finite in this limit.
Hence we may assume $\zeta^{(2)}$ relaxes more rapidly, and 
we may put $d\zeta_{k,p}^{(2)}/dt=0$ in (\ref{eq:eqzeta2}). 
We may then substitute the solution into 
eq. (\ref{eq:eqzeta1}) to obtain
\be
\tau_1\frac{d}{dt}\zeta_{k,p}^{(1)}=
-\frac{3\epsilon\ell}{4}[D_\perp p^2+D_\parallel k^2+Kp^4]\zeta_{k,p}^{(1)}
\label{eq:phaseeq1}
\ee
where 
\be
D_\perp =\frac{4\epsilon}{3\ell}-\frac{8\alpha}{\ell^2}
\sum_{Q}\frac{1}{(Q^2+\kappa^2)^2}(1-\cos 2Qw)
\label{eq:Dperp}
\ee
\begin{eqnarray}
D_\parallel &=&\frac{24\alpha}{\ell^2}
\sum_{Q}\frac{1}{(Q^2+\kappa^2)^2}(1-\cos 2Qw)\nonumber\\
{}&-&\frac{16\alpha}{\sum_{Q}\frac{\cos 2Qw}{Q^2+\kappa^2}}
(\frac{1}{\ell}\sum_{Q}\frac{Q\sin 2Qw}{(Q^2+\kappa^2)^2})^2
\label{eq:Dparallel}
\end{eqnarray}
and
\be
K=\frac{8\alpha}{\ell^2}\sum_{Q}\frac{1}{(Q^2+\kappa^2)^3}(1-\cos 2Qw)
\label{eq:Dk4}
\ee
In eq. (\ref{eq:phaseeq1}), we have factored 
out $3\epsilon\ell/4$. 
The reason will become clear in eq. (\ref{eq:zoubunqqq}) below.
Eq. (\ref{eq:phaseeq1}) with the coefficients given by (\ref{eq:Dperp}),
(\ref{eq:Dparallel}) and (\ref{eq:Dk4}) 
is the general form for layered structures.
The coefficient $K$ is always positive and $D_\parallel$ is positive at least
when $w \approx \ell/4$. Thus the layered
structure in the present model system does not cause any linear
instabilities in this regime. 

The coefficient $D_\perp$ must be identically zero when 
the spatial period $\ell$ is the equilibrium period. This will be verified
separately in the next section.

\section{Determination of the period}
\label{sec:period}
When the system has a free energy functional, the equilibrium period should be
determined by its minimization. In this section we show that this is identical
to the condition $D_\perp =0$. Although this is a well--known fact (see, e.g.,
ref. \cite{KO1}), 
a verification is necessary to confirm the validity of the present approach.

We consider a one--dimensional system with the system size $L(>>\ell)$ and
write the free energy per unit length as
\begin{equation}
\hat{F}=\frac{1}{2L}\int dx\int dx's(x-x')u(x)u(x')+\frac{1}{L}\int dxW(u)
\label{eq:hatFx}
\end{equation}
where $s(x-x')$ is the nonlocal interaction. In the case of (\ref{eq:free})
it is given by
\begin{equation}
s(x-x')=\epsilon^2\frac{\partial^2}{\partial x\partial x'}\delta (x-x')
+\alpha G(x-x')
\label{eq:sxx}
\end{equation}
The second term of (\ref{eq:hatFx}) is the local part. We assume that the
solution of $\delta F/\delta u=0$  is spatially periodic.

Now suppose that the system is expanded uniformly so that the period becomes
$\ell (1+\delta)$ ($\delta <<1$). Thus $u$ and $L$ are transformed as
$u(x) \rightarrow u(x/(1+\delta)), L \rightarrow L(1+\delta)$. The free
energy changes as
\begin{equation}
\hat{F}=\frac{1+\delta}{2L}\int dy\int dy's((y-y')
(1+\delta))u(y)u(y')+\frac{1}{L}\int dyW(u)
\label{eq:hatFy}
\end{equation}
where $x/(1+\delta)=y$. The free energy increase $\triangle \hat{F}$ 
is, therefore,
given up to
$O(\delta)$ by
\begin{eqnarray}
\triangle \hat{F}&=&\frac{\delta}{2L}\int dy\int dy'
s(y-y')u(y)u(y')\nonumber\\
{}&+&\frac{\delta}{2L}\int dy\int dy'
(y-y')[\frac{ds(y-y')}{d(y-y')}]u(y)u(y')\nonumber\\
{}&=&-\frac{\delta}{2L}\int dy\int dy'(y-y')s(y-y')[\frac{du(y)}{dy}]u(y')
\label{eq:zoubun}
\end{eqnarray}
This may be rewritten in terms of the Fourier components as
\begin{equation}
\triangle \hat{F}=-\frac{\delta}{2L}\int 
\frac{dq}{2\pi}q[\frac{ds_q}{dq}]u_q u_{-q}
\label{eq:zoubunq}
\end{equation}
If $u$ is the equilibrium solution, 
$\triangle \hat{F}$ must be identically zero.

When
$s(x-x')$ is given by (\ref{eq:sxx}), $s_q=\epsilon^2q^2+\alpha/(q^2+\kappa^2)$
so that we have
\begin{equation}
\triangle \hat{F}=-\frac{\delta}{L}\int 
\frac{dq}{2\pi}(\epsilon^2q^2-\frac{\alpha q^2}{(q^2+\kappa^2)^2})u_q u_{-q}
\label{eq:zoubunqq}
\end{equation}
Let us calculate (\ref{eq:zoubunqq}) explicitly. First of all, 
we define the equilibrium interfacial energy $\sigma$.
When $f(u)$ is given by (\ref{eq:fu}), it is calculated as
\be
\sigma=\epsilon^2 \int dx (\frac{du}{dx})^2=\frac{2\epsilon}{3}
\label{eq:sigma}
\ee
where the integral domain over $x$ is restricted to the vicinity of
an interface.
Note that the interface width dependence of $\sigma$ is different from
that in ordinary critical phenomena. However, this is only apparent and
is due to the
definition of the free energy (\ref{eq:fgl}) where $\epsilon$ appears
as a coefficient of the gradient term.

By using (\ref{eq:sigma}) we note
\begin{equation}
\frac{\epsilon^2}{L}\int \frac{dq}{2\pi}q^2u_q u_{-q}
=\frac{\epsilon^2}{L}\int dx(\frac{du}{dx})^2=2\frac{\sigma}{\ell}
\label{eq:zoubunqq1}
\end{equation}
The factor of 2 comes from the fact that there are two interfaces in each
domain. 

Since we have assumed that $u=1$ 
for $0<x<2w$ and $u=-1$ for $2w<x<\ell$ away from the interfaces,
we obtain
\begin{equation}
\frac{\alpha}{L}\int \frac{dq}{2\pi}\frac{q^2}{(q^2+\kappa^2)^2})u_q u_{-q}
=\frac{8\alpha}{\ell^2}\sum_{Q}\frac{1}{(Q^2+\kappa^2)^2}(1-\cos 2Qw)
\label{eq:zoubunqq2}
\end{equation}
Putting this together with (\ref{eq:zoubunqq1}) we finally obtain

\begin{equation}
\frac{4\epsilon}{3\ell}-\frac{8\alpha}{\ell^2}
\sum_{Q}\frac{1}{(Q^2+\kappa^2)^2}(1-\cos 2Qw)=0
\label{eq:zoubunqqq}
\end{equation}
This precisely agrees with $D_\perp =0$ given by (\ref{eq:Dperp}).

\section{Conserved system}
\label{sec:con}
In a phase separation of binary mixtures, the order parameter
is a local concentration and is a conserved quantity. The 
time-evolution of a conserved system is modelled by
\be
\frac{\partial u}{\partial t} 
=\nabla^2\frac{\delta F}{\delta u} 
\label{eq:con}
\ee
This equation with the free energy given by (\ref{eq:free}) was introduced
to study a microphase separation of block copolymers \cite{BO}. 

When the long range interaction is absent in (\ref{eq:free}), eq. 
(\ref{eq:con}) is called the Cahn-Hilliard equation for
spinodal decomposition, and the
interface equation of motion has been derived. 
However, an interfacial anaylsis for multi-layered systems given by
(\ref{eq:con}) has not previously been obtained.

First we reformulate the diffusive model (\ref{eq:con}) by introducing 
an auxiliary field $v$.
\be
\tau_1\frac{\partial u(\vec{r})}{\partial t} 
=\epsilon^2\nabla^2 u + f(u)-\alpha\int
d\vec{r}'G(\vec{r},\vec{r}')u(\vec{r}',t)-v+h 
\label{eq:con1}
\ee
\be
\tau_2\frac{\partial v(\vec{r})}{\partial t} 
=\nabla^2 v + \frac{\partial u(\vec{r})}{\partial t} 
\label{eq:con2}
\ee
where the long range interaction has been put in (\ref{eq:con1}).
If the constant $\alpha$ were zero in (\ref{eq:con1}), 
the above set of equations
would be essentially the same as the so--called phase field model for melt
growth
\cite{CG}, \cite{KOBA}. 
In this case, the variable $v$ is the local temperature and the 
last term in (\ref{eq:con2}) stands for the latent heat production
upon crystalization.

Eq. (\ref{eq:con}) can be recovered as follows. We operate $\nabla^2$ on
both side of eq. (\ref{eq:con1}). 
Putting $\tau_2=0$ in eq. (\ref{eq:con2}), we may write $\nabla^2 v$ 
in terms of $u$, and hence (\ref{eq:con1}) becomes a closed equation for $u$.
Taking the limit $\tau_1 \rightarrow 0$, we obtain eq. (\ref{eq:con}).
Thus the expression of the diffusive model as  eqs. (\ref{eq:con1}) and 
(\ref{eq:con2}) enables us to analyze the interface dynamics almost parallel
to the non-conserved case.

However, there are two important differences. One is the fact that
$v_I$ now contains the time derivative of $\zeta^{\pm}$
because of $\partial u/\partial t$ in (\ref{eq:con2}). 
Putting $\tau_1=0$, eq. (\ref{eq:bvp1I}) becomes
\be
0=\epsilon^2H - 3\epsilon(v_I-h)
\label{eq:kaimencon}
\ee
It should be noted, however, that the contribution 
from the third term in eq. (\ref{eq:con1})
is also included in $v_I$. That is, $v_I$ consists of two parts 
$v_I=v_I^{(1)} + v_I^{(2)}$
where $v_I^{(1)}$ comes from the third term in (\ref{eq:con1}) 
and is given by (\ref{eq:vI1p}) and (\ref{eq:vI1m})
whereas $v_I^{(2)}$ comes from the fourth term in (\ref{eq:con1}), 
which will be evaluated below.
Eq. (\ref{eq:con2}) can be solved as
\be
v(\vec{r},t)=\int d\vec{r}'G_c(\vec{r},\vec{r}')\frac{\partial
u(\vec{r}',t)}{\partial t}
\label{eq:vrtcon}
\ee
It should be noted that the Green function in (\ref{eq:vrtcon}) is not the
screened one
(\ref{eq:greenkappa}) but the Coulomb Green function (\ref{eq:green}).
Using the fact that
\be
\frac{\partial u}{\partial t}=-2\sum_{m}[\dot{\zeta}_m^-
\delta(x-m\ell+w-\zeta_m^-(y))-\dot{\zeta}_m^+
\delta(x-m\ell-w-\zeta_m^+(y))]
\label{eq:dudt}
\ee
where the factor 2 comes from the gap of $u$ at the interfaces,
we readily obtain up to linear order in $\zeta^{\pm}$
\be
\hat{v}_I^{(2)+}=\frac{2}{\ell}
\sum_{Q(=0)}\frac{1}{(k-Q)^2 +p^2}
[\dot{\zeta}_{k,p}^+-e^{-2iQw}\dot{\zeta}_{k,p}^-]
\label{eq:vI2p}
\ee

\be
\hat{v}_I^{(2)-}=\frac{2}{\ell}
\sum_{Q(=0)}\frac{1}{(k-Q)^2 +p^2}
[e^{2iQw}\dot{\zeta}_{k,p}^+-\dot{\zeta}_{k,p}^-]
\label{eq:vI2m}
\ee
where $\hat{v}_I^{(2)\pm}$ is the Fourier component of $v_I^{(2)\pm}$.
In these expressions we have emphasized that the $n=0$ component
is included in the summations.

The other difference from the non-conserved case is that we have to take account
of the conservation law in the interface equation of motion.
In the non--conserved case, we consider the zeroth--solution separately as 
(\ref{eq:kvI0}) which determines the ratio $w/\ell$.
However, in a conserved system this ratio is fixed uniquely in the strong
segregation limit by the average volume fraction. 
In fact, when there is no current across
the system boundary, the conservation law requires 
\be
\frac{d}{dt}\sum_n \int dy \zeta_n^{(2)}=0
\label{eq:concon}
\ee
The constant $h$ in eq. (\ref{eq:con1}) plays a role of
a Lagrange multiplier associated with the condition (\ref{eq:concon}).

Substituting (\ref{eq:vI2p}) and (\ref{eq:vI2m}) 
into (\ref{eq:kaimencon}) and retaining the zero--th order term, 
we obtain
\begin{eqnarray}
\frac{6\epsilon}{\ell}\sum_{Q(=0)}\frac{1}{(k-Q)^2 +p^2}
&[&(1-\cos 2Qw)\dot{\zeta}_{k,p}^{(1)}
-i \sin 2Qw \dot{\zeta}_{k,p}^{(2)}]\nonumber\\
{}&=&-[\epsilon^2 p^2+
\frac{6\epsilon\alpha}{\ell}(3k^2-p^2)
\sum_{Q}\frac{1}{(Q^2+\kappa^2)^2}(1-\cos 2Qw)]\zeta_{k,p}^{(1)}\nonumber\\
{}&+&\frac{12i\epsilon\alpha}{\ell}k\zeta_{k,p}^{(2)}
\sum_{Q}\frac{Q}{(Q^2+\kappa^2)^2}\sin 2Qw\nonumber\\
{}&-&\frac{6\epsilon\alpha}{\ell}\sum_{Q}
\frac{1}{(Q^2+\kappa^2)^3}(1-\cos 2Qw)p^4\zeta_{k,p}^{(1)}
\label{eq:eqzetacon1}
\end{eqnarray}

\begin{eqnarray}
\frac{6\epsilon}{\ell}\sum_{Q(=0)}\frac{1}{(k-Q)^2 +p^2}
&[&(1+\cos 2Qw)\dot{\zeta}_{k,p}^{(2)}
+i \sin 2Qw \dot{\zeta}_{k,p}^{(1)}]\nonumber\\
{}&=&-[\epsilon^2p^2+
\frac{12\epsilon\alpha}{\ell}\sum_{Q}
\frac{1}{Q^2+\kappa^2}\cos 2Qw]\zeta_{k,p}^{(2)}\nonumber\\
{}&-&\frac{6\epsilon\alpha}{\ell}(3k^2-p^2)
\sum_{Q}\frac{1}{(Q^2+\kappa^2)^2}(1+\cos 2Qw)\zeta_{k,p}^{(2)}\nonumber\\
{}&-&\frac{12i\epsilon\alpha}{\ell}k\zeta_{k,p}^{(1)}
\sum_{Q}\frac{Q}{(Q^2+\kappa^2)^2}\sin 2Qw \nonumber\\
{}&-&C\delta(k)\delta(p)
\label{eq:eqzetacon2}
\end{eqnarray}
where
\be
C=\frac{2\alpha}{\ell}\sum_{Q}\frac{\sin 2Qw}{Q(Q^2+\kappa^2)} -h
\label{eq:C}
\ee
The right hand sides of (\ref{eq:eqzetacon1}) and (\ref{eq:eqzetacon2})
have been expanded in powers of $k$ and $p$ and are the same as those in
(\ref{eq:eqzeta1}) and (\ref{eq:eqzeta2}) respectively.  
The unknown constant $h$ or, equivalently $C$, 
is determined by (\ref{eq:concon}).
That is, in the limit $k, p \rightarrow 0$ in eq. (\ref{eq:eqzetacon2}) 
we obtain
\be
C=-\frac{12\epsilon\alpha}{\ell}\sum_{Q}\frac{\cos 2Qw}{Q^2+\kappa^2}
\zeta_{0, 0}^{(2)}
\label{eq:CC}
\ee
Note from (\ref{eq:CC}) that 
the "zero--th" order term is actually not strictly zero--th order,
but it turns out to be linearly dependent on $\zeta^{(2)}$.

It should be noted that the summations on the left hand side of 
eqs. (\ref{eq:eqzetacon1}) and (\ref{eq:eqzetacon2}) contain the $Q=0$
component. In particular,  eq. (\ref{eq:eqzetacon2}) produces the factor
$1/(k^2+p^2)$ which originates from the conservation law and 
is dominant in the long wavelength limit. 
Thus eqs. (\ref{eq:eqzetacon1}) and (\ref{eq:eqzetacon2}) are
approximated as
\be
\dot{\zeta}_{k,p}^{(1)} =-\Gamma^{(1)}
[(D_\perp p^2+Dk^2+Kp^4)\zeta_{k,p}^{(1)}
-ikB_2\zeta_{k,p}^{(2)}]
\label{eq:eqzetacon4}
\ee
\be
\dot{\zeta}_{k,p}^{(2)} =-\Gamma^{(2)}(k^2+p^2)
[B_1\zeta_{k,p}^{(2)}+ikB_2\zeta_{k,p}^{(1)}]
\label{eq:eqzetacon3}
\ee
where
\be
D=\frac{24\alpha}{\ell^2}\sum_{Q}\frac{1-\cos 2Qw}{(Q^2+\kappa^2)^2}
\label{eq:D}
\ee
\be
\Gamma^{(1)}= \frac{\ell^2}{8\sum_{Q\ne0}\frac{1-\cos 2Qw}{Q^2}}
\label{eq:Gamma1}
\ee
\be
\Gamma^{(2)}= \frac{\ell^2}{16}
\label{eq:Gamma2}
\ee
\be
B_1=\frac{16\alpha}{\ell^2} \sum_{Q}\frac{\cos 2Qw}{Q^2+\kappa^2}
\label{eq:B1}
\ee
\be
B_2=\frac{16\alpha}{\ell^2}\sum_{Q}\frac{Q\sin 2Qw}{(Q^2+\kappa^2)^2}
\label{eq:B2}
\ee
We have omitted the C term, and $D_\perp \rm and\ K$ are given by
eqs. (\ref{eq:Dperp}) and (\ref{eq:Dk4}), respectively.
The factor $\Gamma^{(2)}$ has been factored out such that the $B_2$
term in (\ref{eq:eqzetacon4}) and (\ref{eq:eqzetacon3}) have
the same coefficient.
In this way we have a coupled set of equations for $\zeta^{(1)}$ 
and $\zeta^{(2)}$, whose relaxation rates vanish in the limit
$k, p \rightarrow 0$.

There are two cases where the relaxation of $\zeta^{(2)}$
is much faster than that of $\zeta^{(1)}$. One is the limit
$\kappa \rightarrow 0$ where $B_1$ becomes very large because of
the $Q=0$ component. Note that $D$ remains finite.
The other is the case that 
$D_\perp =0$ and $k^2 <<p^2$. In these situations we
can eliminate $\zeta^{(2)}$ by putting
$\dot{\zeta}^{(2)}=0$:
\be
\frac{d}{dt}\zeta_{k,p}^{(1)}=
-\Gamma^{(1)}[D_\parallel k^2+Kp^4]\zeta_{k,p}^{(1)}
\label{eq:phaseeqcon1}
\ee
where $D_\parallel$ is the same as in 
(\ref{eq:Dparallel}). 

\section{Binary fluids}
\label{sec:fluid}
Equation (\ref{eq:con}) has been applied to a microphase separation of 
diblock copolymers. When the sizes of the 
two blocks are nearly equal, the system
exhibits a lamellar structure in the microphase separated state.
The kinetics of this phase separation have been studied mainly by omitting 
any hydrodynamic effects. However, in some situations such 
as when the system is subjected to shear flow, hydrodynamic effects are known
to cause
an instability of a lamellar structure \cite{Fred}. 

Another example in fluids is the formation of a periodic structure
within a
Langmuir monolayer \cite{ABJ}, \cite{MM}. However, since this is an
adsorbed system
between an air/water interface, 
the hydrodynamics are different from that in bulk \cite{KATZ}. One needs a
separate treatment
of its interface dynamics which we do not enter here.

In this section, we consider the conserved system coupled with the local
velocity field $\vec{v}$.
\be
\frac{\partial u(\vec{r})}{\partial t} +\vec{v}\cdot \vec{\nabla}u
=\nabla^2[-\epsilon^2\nabla^2 u - f(u)+\phi] 
\label{eq:bf1}
\ee
\be
\frac{\partial \vec{v}}{\partial t} 
=\eta \nabla^2 \vec{v} + (\vec{\nabla}u)[-\epsilon^2\nabla^2 u - f(u)+\phi]
\label{eq:bf2}
\ee
\be
\nabla^2 \phi -\kappa^2 \phi+ \alpha u=0
\label{eq:bf3}
\ee
where $\eta$ is the shear viscosity. 
In order to avoid confusion of the velocity field $\vec{v}$ with 
the auxiliary field $v$, we have
introduced the notation $\phi$ instead of $v$ for the auxiliary density.
We have imposed the incompressibility condition
$\vec{\nabla}\cdot\vec{v}=0$ and eliminated the pressure term.
Thus $\vec{v}$ has only transverse components.
The nonlinear convective term $\vec{v}\cdot \nabla \vec{v}$ 
has been ignored in eq. (\ref{eq:bf2}).

A general form of the interface equation can be derived as follows.
We temporarily ignore the diffusive 
coupling of the order parameter given by the right
hand side of (\ref{eq:bf1}) and consider only 
the hydrodynamic coupling $\vec{v}\cdot \vec{\nabla}u$.
At the final stage we will take account of the diffusive effect
given by the right hand side of eq. (\ref{eq:bf1}).
Thus the interface velocity $V$ is given from (\ref{eq:bf1}) by
\be
V=-\vec{v}_I\cdot \vec{n}
\label{eq:bfV}
\ee
where the unit normal vector $\vec{n}$ is directed inward to a $u=1$ domain
and $\vec{v}_I$ is the fluid velocity at the interface.
Putting $\partial \vec{v}/ \partial t=0$, 
as usual assuming velocity fields relax rapidly compared to interface motion,
we can solve eq. (\ref{eq:bf2}).
Substituting it into (\ref{eq:bfV}), we obtain
\be
V(a)=\int da' \sum_{\alpha, \beta}n^{\alpha}(a)T^{\alpha \beta}(a, a')
n^{\beta}(a')[\sigma H(a')-2\alpha \phi_I(a')]
\label{eq:bfkaimen}
\ee
where $a$ and $a'$ stand for the position on the interface. The tensor
$T^{\alpha \beta}$ is the Oseen tensor whose Fourier transform is given by
\be
\hat{T}_q^{\alpha \beta}=\frac{1}{\eta q^2}[\delta_{\alpha, \beta}
-\frac{q^\alpha q^\beta}{q^2}]
\label{eq:Oseen}
\ee
The constant $\sigma$ is the interfacial energy defined by (\ref{eq:sigma}).
The value $\phi_I$ is the same as $v_I$ given by (\ref{eq:vI1p}) and
(\ref{eq:vI1m}).
In this case the zero--th order constant part of $\phi_I$ does not 
contribute to (\ref{eq:bfkaimen}) because of the relation
\be
\int da' \sum_{\alpha, \beta}n^{\alpha}(a)T^{\alpha \beta}(a, a')
n^{\beta}(a')=0
\label{eq:T0}
\ee
Hence we may consider only the first order corrections given by 
(\ref{eq:vI1p}) and (\ref{eq:vI1m}) for $\phi_I$ in (\ref{eq:bfkaimen}).

The interface equation of motion can be derived from (\ref{eq:bfkaimen}).
Note that $\vec{n}$ has only the x--component to lowest order
and $n^x=-1$ for a right interface and $n^x=1$ for a left interface.
Using the facts that $V=\pm \partial \zeta^{\pm}/\partial t$ and
$H=\pm \partial^2 \zeta^{\pm}/\partial y^2$,
we obtain
\begin{eqnarray}
\frac{d}{dt}\vec{\zeta}&=&\frac{2}{3\epsilon\ell}
\sum_{Q(=0)}T_{k-Q, p}^{xx}N(Q)
[-\epsilon^2p^2+\frac{6\epsilon\alpha}{\ell}
\sum_{Q'}\frac{1}{Q'^2+\kappa^2}(1-\cos 2Q'w)\nonumber\\
{}&-&\frac{6\epsilon\alpha}{\ell}\sum_{Q'}\frac{1}{(k-Q')^2+p^2+\kappa^2}
M(Q')]\vec{\zeta}
\label{eq:eqzetafluid}
\end{eqnarray}
where $\vec{\zeta}$ is a column vector with $\zeta_1=\zeta_{k,p}^{(1)}$ and
$\zeta_2=\zeta_{k,p}^{(2)}$ and $M(Q)$ is a 2$\times$2 matrix with the 
components $M_{11}=1-\cos2Qw$, $M_{12}=-i\sin2Qw$, $M_{21}=i\sin2Qw$ and
$M_{22}=1+\cos2Qw$.  The 2$\times$2 matrix $N(Q)$ has
the components $N_{11}=1+\cos2Qw$, $N_{12}=i\sin2Qw$, $N_{21}=-i\sin2Qw$ and
$N_{22}=1-\cos2Qw$.

Now we expand (\ref{eq:eqzetafluid}) in powers of $k$ and $p$
where it should be noted that the summation over $Q$ contains the $Q=0$
component.
Using the relations
\be
T_{k, p}^{xx}=\frac{p^2}{\eta(k^2+p^2)^2}
\label{eq:Txx0}
\ee
\be
T_{k-Q, p}^{xx}\approx \frac{p^2}{\eta Q^4}
\label{eq:Txx}
\ee
eq. (\ref{eq:eqzetafluid}) becomes in the long wavelength limit
\be
\dot{\zeta}_{k,p}^{(1)} =-\frac{p^2}{\eta(k^2+p^2)^2}
[(D_\perp p^2+Dk^2+Kp^4)\zeta_{k,p}^{(1)}
-ikB_2\zeta_{k,p}^{(2)}]
\label{eq:eqzetafluid1}
\ee
\be
\dot{\zeta}_{k,p}^{(2)} =-\frac{\gamma p^2}{\eta}
[B_1\zeta_{k,p}^{(2)}+ikB_2\zeta_{k,p}^{(1)}]
\label{eq:eqzetafluid2}
\ee
where
\be
\gamma  =\frac{1}{2}\sum_{Q\ne0}\frac{1-\cos2Qw}{Q^4}
\label{eq:smallgamma}
\ee
The reason for the appearance of the factor $p^2$ in (\ref{eq:eqzetafluid2})
is clear. We have so far ignored the diffusive effect in (\ref{eq:bf1}) and
considered only the hydrodynamic interactions. Therefore, the concentration
must be conserved within each domain, and hence the effective diffusion 
exists only
along the y--axis in (\ref{eq:eqzetafluid2}), i.e., along the interface.

By comparing (\ref{eq:eqzetafluid1}) and (\ref{eq:eqzetafluid2}), we note that
if $k^2<<p^2$ and $K<<\gamma B_1$ for $D_\perp =0$, 
the relaxation of $\zeta^{(2)}$ is rapid 
so that we can eliminate it to obtain
\be
\frac{d}{dt}\zeta_{k,p}^{(1)}=
-\frac{p^2}{\eta(k^2+p^2)^2}[D_\parallel k^2+Kp^4]\zeta_{k,p}^{(1)}
\label{eq:phaseeqfluid}
\ee
This is also justified when $\kappa \rightarrow 0$ since the constant $B_1$ 
becomes infinite in this limit.
The coefficients are the same as those in the previous sections.
If necessary, we may add to (\ref{eq:phaseeqfluid}) the contribution 
from the diffusive effect given by eq. (\ref{eq:phaseeqcon1}).
This equation has been obtained by several authors \cite{MIL}, \cite{ONUKI}.
However, under general conditions, the phase variable $\zeta^{(1)}$ has to be
coupled with the other slow mode $\zeta^{(2)}$ as in the set
eq. (\ref{eq:eqzetafluid1})
with eq. (\ref{eq:eqzetafluid2}).

\section{Discussion}
\label{sec:disc}
We have presented a systematic method to develop the equations
governing interface dynamics
in layered structures. It is emphasized that the basic equations
(\ref{eq:bvp1}), (\ref{eq:bvp2}) and (\ref{eq:con1}),
(\ref{eq:con2}) do not possess any Lyapunov functional when $\tau_1$
and $\tau_2$ are finite. Although we have taken
a limit in which these parameters vanish, this is simply 
because we have demonstrated
our theory mainly for the systems near equilibrium. 
The interface dynamics can be formulated
without taking these limits as has already been shown for 
a nonconserved reaction--diffusion system in Ref. \cite{OIT}. 
In the present paper, we have extended the methods 
to conserved systems and binary fluids. 

The motion of domains can be divided into two parts. One is the collective
motion of center of gravity of each domain, which is described by a phase 
variable $\zeta^{(1)}$. The other is the internal motion of each 
domain denoted by $\zeta^{(2)}$. 

Equations of motion for $\zeta^{(1)}$ and $\zeta^{(2)}$
in conserved and non--conserved systems take the following 
form:
\begin{equation}
\frac{\partial \zeta^{(1)}}{\partial t}=
-\Gamma^{(1)}\frac{\delta H}{\delta \zeta^{(1)}}
\label{eq:phaseeqd1}
\end{equation}
\begin{equation}
\frac{\partial \zeta^{(2)}}{\partial t}=
-\Gamma^{(2)}(i\vec{\nabla})^a \frac{\delta H}{\delta \zeta^{(2)}}
\label{eq:phaseeqd2}
\end{equation}
where $a=0$ for a non--conserved case and $a=2$ for a conserved case
and $H$ (which should not be confused with the mean curvature defined by
(\ref{eq:mean})) is given by
\begin{equation}
H=\frac{1}{2}\int dxdy [D(\frac{\partial \zeta^{(1)}}{\partial x})^2
+D_\perp (\frac{\partial \zeta^{(1)}}{\partial y})^2
+K(\frac{\partial^2 \zeta^{(1)}}{\partial y^2})^2
+B_1(\zeta^{(2)})^2 + 2B_2 \frac{\partial \zeta^{(1)}}{\partial x} \zeta^{(2)}]
\label{eq:phaseH}
\end{equation}
This Hamiltonian is invariant under the transformation $x \rightarrow -x$
since $\zeta^{(1)}$ changes its sign whereas $\zeta^{(2)}$ is invariant.
Using the same reasoning, the coupling term
$(\partial \zeta^{(1)}/\partial y)\zeta^{(2)}$ should not exist.
In the case of binary fluids, we can write the equations formally as
\begin{equation}
\frac{\partial \zeta^{(1)}}{\partial t}=
\frac{1}{\eta(\nabla^2)^2}\frac{\partial^2}{\partial y^2}
\frac{\delta H}{\delta \zeta^{(1)}}
\label{eq:phasefluid1}
\end{equation}
\begin{equation}
\frac{\partial \zeta^{(2)}}{\partial t}=
\frac{\gamma}{\eta}\frac{\partial^2}{\partial y^2}
\frac{\delta H}{\delta \zeta^{(2)}}
\label{eq:phasefluid2}
\end{equation}
These sets of equations show that although $\zeta^{(2)}$ has a ``mass''
term with 
coefficient $B_1$ in $H$, it does not produce a finite relaxation rate
of $\zeta^{(2)}$ in the long wavelength limit except 
in the non-conserved case. 
In the conserved system and binary fluids, the variable $\zeta^{(2)}$ is
generally
a slow mode in the limit of long wavelength deformations because of 
the conservation law.
As we have remarked, $\zeta^{(2)}$ can be eliminated only when the
deformation normal to the layers
is spatially sufficiently weak compared with the undulation of the layers.
In fact, eqs. (\ref{eq:phasefluid1}) and (\ref{eq:phasefluid2}) give us
the relaxation rate $\Omega^{(2)}$ of $\zeta^{(2)}$ approximately as 
\begin{equation}
\Omega^{(2)}=\frac{\gamma p^2 B_1}{\eta}
[1-\frac{B_2^2}{\gamma B_1^2}\frac{k^2}{(k^2 + p^2)^2}]
\label{eq:Omega2}
\end{equation}
where the second term is the correction due to 
the coupling with $\zeta^{(1)}$, which
is, however, not necessarily small for $p \rightarrow 0$.

Note that the case of pure Coulomb long range interactions has to
be handled carefully with the limit $\kappa \rightarrow 0$. In all
cases studied here, when $\kappa$ is sufficiently small, the 
internal mode $\zeta^{(2)}$ can be eliminated adiabatically,
and the phase variable description involving $\zeta^{(1)}$ is
complete.

In this paper, we have restricted ourselves 
to a one--dimensional layered structure.
However, the theory is easily extended to those in higher dimensions
such as a hexagonal structure of disk--shaped domains and a 
face centered cubic structure of spherical
domains. Evaluation of nonlinear terms is also straightforward although
the calculations are more involved.

\vspace{0.2in}

\noindent {\bf Acknowledgments:}
TO is grateful to Masao Doi, Ayako Tetsuka and Yumino Hayase 
for valuable discussions at the
early stage of this work.
This work was supported by the Grand--in--Aid of Ministry of Education,
Science and Culture of Japan.
DJ is grateful for the support of the NSF under DMR92-17935.
\vspace{0.2in}

\begin{center}
{\bf APPENDIX}
\end{center}
\vspace{0.2in}

Here we summarize some of the steps in the derivation of the coupled set of
interface equations of motion in section 3. From eqs. (\ref{eq:vrtm})
and (\ref{eq:vIn}), the value of $v$ at an interface is given explicitly by 
\begin{eqnarray}
v_I^{\pm}&=&\alpha\int \frac{d\vec{q}}{(2\pi)^2}\int dy'
\sum_{m}[\int_{b_m^-}^{b_m^+}dx'\frac{1}{q^2+\kappa^2}
\exp(iq_x(b_n^{\pm}-x')+iq_y(y-y'))\nonumber\\
{}&-&\int_{b_m^+}^{b_{m+1}^-}dx'\frac{1}{q^2+\kappa^2}
\exp(iq_x(b_n^{\pm}-x')+iq_y(y-y'))]
\label{eq:a1}
\end{eqnarray}
Note that $b_n^{\pm}$ contains the deformations $\zeta_n^{\pm}(y)$
as in eq. (\ref{eq:vrtpm}).

The zero-th order $v_I^{(0)\pm}$ is obtained by ignoring $\zeta_n^{\pm}$.
From (\ref{eq:a1}) it is given by
\begin{eqnarray}
v_I^{(0)\pm}&=&\alpha\int \frac{d\vec{q}}{(2\pi)^2}\int dy'
\sum_{m}[\int_{m\ell-w}^{m\ell+w}dx'\frac{1}{q^2+\kappa^2}
\exp(iq_x(n\ell\pm w)-iq_xx'+iq_y(y-y'))\nonumber\\
{}&-&\int_{m\ell+w}^{(m+1)\ell-w}dx'\frac{1}{q^2+\kappa^2}
\exp(iq_x(n\ell\pm w)-iq_xx'+iq_y(y-y'))]\nonumber\\
{}&=&\alpha\int \frac{dq_x}{2\pi}
\sum_{m}[\int_{m\ell-w}^{m\ell+w}dx'\frac{1}{q_x^2+\kappa^2}
\exp(iq_x(n\ell\pm w)-iq_xx')\nonumber\\
{}&-&\int_{m\ell+w}^{(m+1)\ell-w}dx'\frac{1}{q_x^2+\kappa^2}
\exp(iq_x(n\ell\pm w)-iq_xx')]
\label{eq:a2}
\end{eqnarray}
The integral over $x'$ in (\ref{eq:a2}) is readily carried out.
By using the Poisson summation formula (\ref{eq:poissonf}),
one can obtain the zero-th solution (\ref{eq:vI0}). 

In order to calculate the first order correction $v_I^{(1)\pm}$, one needs
to expand $\zeta_m^{\pm}$ and $\zeta_n^{\pm}$ contained 
in $b_m^{\pm}$ and $b_n^{\pm}$, respectively, in (\ref{eq:a1}).
Hence the correction consists of two parts
\begin{eqnarray}
v_I^{(1)\pm}&=&\alpha\zeta_n^{\pm}(y)\int \frac{dq_x}{2\pi}
\sum_{m}[\int_{m\ell-w}^{m\ell+w}dx'\frac{iq_x}{q_x^2+\kappa^2}
\exp(iq_x(n\ell\pm w)-iq_xx')\nonumber\\
{}&-&\int_{m\ell+w}^{(m+1)\ell-w}dx'\frac{iq_x}{q_x^2+\kappa^2}
\exp(iq_x(n\ell\pm w)-iq_xx')]\nonumber\\
{}&+&\alpha\int \frac{d\vec{q}}{(2\pi)^2}\int dy'
\sum_{m}\frac{1}{q^2+\kappa^2}
\exp(iq_x(n\ell\pm w-m\ell)+iq_y(y-y'))\nonumber\\
{}&\times&[\zeta_m^+(y')e^{-iq_xw}-\zeta_m^-(y')e^{iq_xw}
-\zeta_{m+1}^-(y')e^{-iq_x(\ell-w)}+\zeta_m^+(y')e^{-iq_xw}]
\label{eq:a3}
\end{eqnarray}
Evaluation of the term which contains the factor $\zeta_n^{\pm}(y)$
is straightforward, which gives us the first term in (\ref{eq:vI1p})
and (\ref{eq:vI1m}). The integrals over $q_y$ and $y'$ in the remaining term
of (\ref{eq:a3}) can be easily carried out by introducing the Fourier transform
(\ref{eq:zetaK}). Using the formula (\ref{eq:poissonf}), one finally obtains the
second term in (\ref{eq:vI1p})
and (\ref{eq:vI1m}).

\vspace{1.0in}
\input epsf
\begin{figure}[h]
\centering
\epsfxsize=5.0in  
\hspace*{0in}  
\epsffile{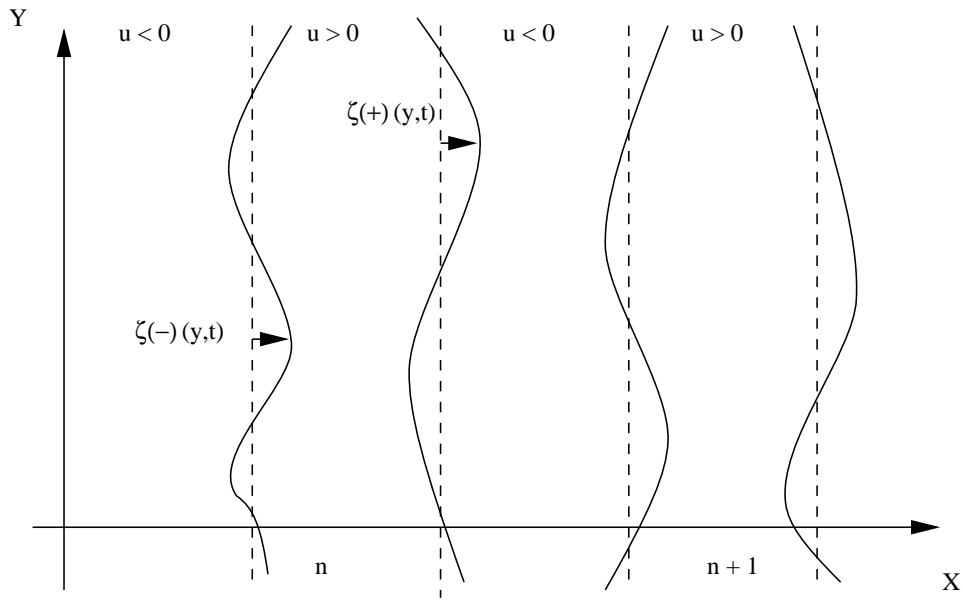} 
\caption{Gently deformed layered structure.
The dotted lines indicate an equilibrium configuration with $w=\ell/2$.}
\label{} 
\end{figure}



\end{document}